\documentclass[twocolumn,aps,amsfonts]{revtex4}
\usepackage{epsfig}
\usepackage{psfig}
\usepackage{bm}
\newcommand{\be}{\begin{equation}}
\newcommand{\ee}{\end{equation}}
\newcommand{\bea}{\begin{eqnarray}}
\newcommand{\eea}{\end{eqnarray}}

\newcommand{\ve}{{\mathbf e}}

\begin{document}

\title{MOND habitats within the solar system}

\author{Jacob Bekenstein$^1$ and Jo\~{a}o Magueijo$^{2,3,4}$ }

\affiliation{$^1$Racah Institute of Physics, Hebrew University of
Jerusalem, Jerusalem 91904, Israel\\
$^2$Perimeter Institute for Theoretical Physics, 31 Caroline St.
N., Waterloo, N2L 2Y5, Canada,\\
$^3$ Canadian Institute for Theoretical Astrophysics,
60 St. George St, Toronto, M5S 3H8, Canada\\
$^4$ Theoretical Physics Group, Imperial College, Prince Consort
Road, London SW7 2BZ, UK}

\begin{abstract}
MOdified Newtonian Dynamics (MOND) is an interesting alternative
to dark matter in extragalactic systems.  We here examine the
possibility that mild or even strong MOND behavior may become
evident well inside the solar system, in particular near  saddle
points of the total gravitational potential.    Whereas in
Newtonian theory tidal stresses are finite at saddle points, they are expected to diverge in MOND, and to remain distinctly large inside a sizeable oblate ellipsoid  around the saddle point.  We work out the MOND effects using the nonrelativistic limit of the T$e$V$e$S theory,  both in  the perturbative nearly Newtonian regime and in the deep MOND regime.  While strong MOND  behavior would be a spectacular ``backyard'' vindication of the theory, pinpointing the MOND-bubbles in the setting of the
realistic solar system  may be difficult.  Space missions, such as the LISA Pathfinder, equipped with sensitive accelerometers, may be able to explore the larger perturbative region.
\end{abstract}

\pacs{PACS Numbers: 1,2,2 }


\date{\today}

\maketitle

\section{Introduction}

MOND~\cite{milg1} is a scheme for explaining extragalactic
phenomenology without invoking dark matter.  It has been very
successful in this job despite its rather rudimentary
form~\cite{sanmcg}. In the Lagrangian formulation of
MOND~\cite{aqual} the physical gravitational potential $\Phi$,
which gives test particle acceleration by $\mathbf
{a}=-\nabla\Phi$, is determined by the modified Poisson equation
\be               \nabla\cdot[\tilde\mu(|\nabla\Phi |/a_0)\nabla\Phi]=4\pi
G\tilde\rho, \label{AQUAL} \ee
where $\tilde\rho$ is the baryonic
mass density, $a_0\approx 10^{-10}$ m s$^{-2}$ is Milgrom's
characteristic acceleration, and the function $\tilde\mu(x)$ is required to approximate its argument for $x\ll 1$ and to approach unity for $x\gg 1$.  The form \be              \tilde\mu(x)=x(1+x)^{-1}\label{tildemu1} \ee             has been quite successful in modelling galaxy rotation curves  without invoking dark matter.  The theory encapsulated in Eq.~(\ref{AQUAL}) has recently been reformulated as a consistent covariant gravitation theory named T$e$V$e$S~\cite{teves}.  Alternatives to this theory have been considered~\cite{BSTV,Aegean,many}, but they shall not be employed in the present work.

Are MOND effects of importance in the solar system (henceforth SS)?  Milgrom was the first to consider the effects of MOND on the properties of long period comets originating in the Oort cloud~\cite{milg1}.  It was later observed that relativistic theories with a MOND limit can easily predict anomalously large perihelion precessions of the planets~\cite{can,Aegean}.  With the discovery of the ``Pioneer anomaly''~\cite{pioneer} much speculation was directed towards a possible MONDian origin of the effect~\cite{Hellemans}.  Relativistic theories with MOND phenomenology tend to produce a radial drift of the Kepler constant in the SS in the same sense as would correspond to the claimed Pioneer effect, though not always of the claimed magnitude~\cite{sanmcg,teves,Aegean,M}.   The persistence of Pioneer-type effects in a variety of scalar-tensor theories of MOND, and the hurdles faced by such theories from precision SS tests has been emphasized by Sanders~\cite{Aegean}.

In this paper we search for other sites deep inside the SS where strong MOND behavior might put the MOND phenomenon at the reach of spacecraft measurements. Strong MOND behavior is triggered by a low gradient in the {\it total} Newtonian potential $\Phi_N$ (the deep MOND regime is that where $|\nabla\Phi|\ll a_0$). Two apparent candidates for strong MOND regions fail this criterion. Most obviously we have gravitational perturbations, such as those accounting for the non-relativistic component of the perihelion of Mercury precession, or Neptune's influence upon Uranus' orbit. Most of these have a very low potential gradient, and would by themselves be in the MOND regime. However the gradient of the total $\Phi_N$is not small, so their effect falls in the Newtonian regime.  Issues like the stability of the SS or its detailed dynamics are not expected to be appreciably different in MOND.

The Lagrange points are another apparent possibility for strong MOND regions. They are the five stationary points of the two-body dynamics; for example L1 is the point between the  Earth and the Sun where a
test mass would be in inertial motion,  moving neither towards the Sun nor the Earth. Each Lagrangian point orbits  the Sun  with the same frequency as the Earth, so the gradient of $\Phi_N$ at it must  cancel the corresponding centrifugal acceleration, and is thus not especially small. This does not mean, as we shall see, that perturbative effects around these points are not present; however strong MONDian behavior is certainly not expected.

By contrast,  the saddle (or extremum) point (henceforth SP) of $\Phi_N$ between two gravitating bodies is evidently in the deep MOND regime, since $\nabla\Phi_N=0$ there.  One such point exists between any two gravitating bodies, potentially providing a testing ground for strong MONDian behavior. SPs are not inertial, but may be visited by free-falling test bodies; they are encased by small ``bubbles'' within which strong MOND effects are expected.

In what follows  we study the structure of the gravitational field in increasingly smaller neighbourhoods of the  SP for a binary mass system.  Principally we look at the encasing Newtonian region, at the enclosed quasi-Newtonian sector and at the deep MOND region at the heart of the bubble.  The strongly nonlinear character of the MOND phenomenon makes the analysis complicated, and a variety of analytical as well as numerical strategies have been utilized.

The plan of this paper is as follows. In Sec.~\ref{formal} we lay down the  framework for dealing with the two (and  many) body problem in MOND. In Sec.~\ref{bubbles} we use heuristic arguments to find the location and extent of the principal MOND bubble for a binary system.  One of the approximations thus made is replaced by a more exact treatment in Sec.~\ref{curlsec}. In Sec.~\ref{quasi} we perturbatively calculate the gravitational field in the quasi-Newtonian region still far from the SP point. In Sec.~\ref{extreme} a combination of numerical and analytical approaches is used to deduce the gravitational field very near the SP where MOND is dominant. In Section~\ref{total} se discuss a number of complications to the above stylized treatment that arise from the many-body nature of the real SS. The prospect of a direct test using the LISA Pathfinder   project~\cite{lisapf} is briefly discussed in Sec.~\ref{LPF}.  Issues connected with the behavior of gravity in the spacecraft's frame are elucidated in Sec.~\ref{capsule}.  We conclude in Sec.~\ref{concs} with a statement on how our work might help settling the controversy between dark matter and MOND.

\section{The framework of the two-body problem}\label{formal}

Where needed we use units with $c=1$.    We base the analysis on the non-relativistic limit of T$e$V$e$S~\cite{teves};  AQUAL~\cite{aqual} and other Lagrangian formulations of MOND have similar form and many of our conclusions may apply to all of them.
In T$e$V$e$S the MOND behaviour is driven by a dynamical (and dimensionless) scalar field $\phi$ such that the physical potential $\Phi$ in which a body falls is given by  $\Phi=\Phi_N +\phi$, where $\Phi_N$ is the usual Newtonian potential
(inferred from the metric component $g_{00}$). In the non-relativistic
regime $\phi$ is governed by the equation
\be              \label{phieq}
\nabla\cdot [\mu(kl^2 (\nabla\phi)^2)\nabla \phi]=kG{\tilde \rho}
\ee
where $k$ is a coupling constant and $l$ is a length scale which determines the Milgrom acceleration by
\be
a_0={\sqrt {3 k}\over 4\pi l}\approx 10^{-10} {\rm m\ s}^{-2}
\ee
(we are setting $\Xi$, as defined in~\cite{teves}, to unity; thus we ignore the slight renormalization of the gravitational constant in  T$e$V$e$S so that here $G_N= G$).  In Eq.~(\ref{phieq})   $\mu$ is a free function not to be confused with Milgrom's $\tilde\mu$.  Ref.~\cite{teves} proposed a particular form for it. The deep MOND regime is signalled by the low gradient of the scalar field $\phi$; in this regime
\be
 \label{mumond}
\mu\approx {k\over 4\pi}{|\nabla\phi|\over a_0}.
\ee
For strong gradients the $\mu$ proposed in Ref.~\cite{teves} grows crudely as $(|\nabla\phi|/a_0)^{2/3}$.  This has the effect of suppressing the contribution of $\nabla\phi$ to $\nabla\Phi$  thus bringing in the Newtonian regime.  In spherically symmetric systems T$e$V$e$S with any $\mu$ satisfying Eq.~(\ref{mumond}) goes over into the Lagrangian MOND theory (\ref{AQUAL}) with Milgrom's $\tilde\mu$ given by $\tilde\mu=(1+k/4\pi\mu)^{-1}$.  Although this point is not well explored, it is quite possible that in less symmetric systems T$e$V$e$S does not go over to an exactly MOND behavior.  For this reason we base this paper on the non-relativistic limit of T$e$V$e$S, and not on Lagrangian MOND.

We need to solve Eq.~(\ref{phieq}) for a two-body source, but that equation is non-linear so the $\phi$ fields due to each body do not superpose. However any non-linear equation  may be \emph{formally} linearized by an appropriate change of variables.  Here this is \be              \mathbf {u}=-{4\pi\mu\over k} \nabla \phi\label{u}\ee              (see~\cite{milg2}, where this technique was first suggested).   We may then add the $\mathbf{u}$ due to each source (which is the Newtonian acceleration) and invert the {\it total} $\mathbf{u}$ at a given point to find $\nabla\phi$.  It is essential that the sum of all sources be performed before inverting to find $\nabla \phi$.

This algorithm  may be applied to any number of components.  But note   that even if a term in the sum is in the  MOND regime, the overall system is not,  unless the total $|\mathbf{ u}|$ is much smaller than $a_0$. (It is because of this feature that the gravitational perturbations in the SS are non-MONDian.)  However, it is also possible to have two components with fields not in the MOND regime such that their common field is MONDian in some region.  Examples are the SPs in the  gravitational potential of two bodies to be studied in this paper.

The only complication with the above technique is that $\mathbf{u}$ is generally not curl-free; indeed it is rather the vector $\mathbf{u}/\mu $ which is curl-free. Thus the full set of equations for $\mathbf{u}$ is
\bea\nabla \cdot \mathbf{u}&=&-4\pi G{\tilde \rho}\label{first}\\\nabla\wedge {\mathbf{u}\over \mu}&=&0. \label{second}
\eea
The first equation tells us that $\mathbf{u}$ equals the Newtonian acceleration $\mathbf{F}^{(N)}=-\nabla\Phi_N$ up to a curl, that is, there must exist a vector field $\mathbf{h}$ such that \be              \label{ufirst} \mathbf{u}=\mathbf{F}^{(N)} +\nabla\wedge \mathbf{h}. \ee              The second equation fixes the  $\mathbf{h}$ (up to a gradient). This operation can only be performed upon the total $\mathbf{u}$, once again stressing the intrinsic non-linearity of the theory. It can be shown that the curl term vanishes in a spherically symmetric situation, or in the quasi-Newtonian regime far away from the source~\cite{aqual,teves}.   Near the SPs neither of these conditions is satisfied and we have to evaluate $\nabla\wedge \mathbf{h}$. However, before plunging into the full problem, let us provide some orientation

\section{Heuristics of the MOND bubbles}
\label{bubbles}

Consider two bodies at distance $R$ with masses $M$ and $m$, with $M\gg m$, so that the system's center of mass may be taken to coincide with the heavier body. To be definite we call them the Sun and the Earth, but we shall explore other couples later. Along the line linking them (the $z$ axis), the Newtonian acceleration is
\be              \mathbf{F}^{(N)}={\left(-{GM\over \tilde r^2}+{Gm\over (R-\tilde r)^2}\right)}\mathbf{ e}_z\ee
where $\tilde r$ is the distance from the Sun and $\mathbf{ e}_z$ is the unit vector in the direction Sun to Earth. The SP of the Newtonian potential $\Phi_N$ resides where $\mathbf{F}^{(N)}=0$, i.e. at
\be              \label{rs} \tilde r=r_s\approx R\left(1-{\sqrt{m\over
M}}\right). \ee

Around this point $\mathbf{F}^{(N)} $ increases linearly as it passes through zero, that is\be              \mathbf{F}^{(N)} \approx A(\tilde r-r_s)\mathbf{e}_z,\label{newton}\ee
where
\be
A=2{GM\over r_s^3}\left(1+\sqrt{{M\over m}}\right)
\label{A}
\ee
is the tidal stress at the SP along the Sun-Earth direction. The full tidal stress matrix is easy to compute.  Let us use cylindrical coordinates centered at the SP,  with the $z$-axis pointing along the Sun-Earth direction, so that we have  $\partial F^{(N)}_z/\partial z = A$ and $\partial F^{(N)}_\varrho/\partial z=0$. From the further condition that the divergence must be zero (outside the Sun and Earth) we have\be              \mathbf{F}^{(N)}=A\big(z\mathbf{e}_z-{\scriptstyle 1\over \scriptstyle 2}\varrho\, \mathbf{e}_\varrho\big).\label{newtonian}\ee

The region around the SP is obviously in the deep MOND regime ($|\mathbf{F}^{(N)}| \ll a_0$).  Thus regardless of the model adopted for $\mu$, we have just on the basis of Eq.~(\ref{mumond}) that\be              \mathbf{ u}\approx -{|\nabla\phi|\over a_0}\nabla\phi=\mathbf{F}^{(N)}  +\nabla\wedge \mathbf{ h}.\label{Ufncurlh}\ee    We may estimate the physical acceleration acting on an object by ignoring the curl term here.  We see that $|\nabla\phi|\ll |\mathbf{F}^{(N)}|$, so that\be              \mathbf{ F}=-\nabla\Phi\approx -\nabla \phi={\sqrt {A a_0}}{ z\mathbf{ e}_z-{\varrho\over 2}\mathbf{ e}_\varrho\over{\left(z^2+{\varrho^2\over 4}\right)}^{1/4} }\,.\label{approx}\ee

In contrast to the Newtonian theory, the tidal stresses here diverge at the SP. This may be understood by applying the rule of thumb that in the deep MOND regime the  square root of the Newtonian acceleration gives the physical acceleration.  According to Eq.~(\ref{newton})  Newtonian acceleration increases linearly along the line Sun-Earth, so the physical acceleration in the deep MOND regime is of the form  $\pm{\sqrt {Aa_0 |\tilde r-r_s|}}$, which has infinite derivative at $r_s$.

The tidal  stresses are expected to remain anomalously high, and gravity to remain in the deep MOND regime (with $|\nabla\phi|\gg |\nabla\Phi_N|$), within a very small oblate ellipsoidal region around the SP. The size of it can be estimated from the condition $|\mathbf{F}^{(N)}|=a_0$ which translates into a major semi-axis (in the $\varrho$ direction) of size
 \be              \label{deltara0}
 \delta \tilde r={2 a_0 \over A} \approx {a_0\over a_m}{\sqrt{m\over M }}R,
 \ee
 where $a_m$ is the acceleration of the smaller mass $m$.   The semi-minor axis (aligned with the $z$ axis), has half this size. For the Sun-Earth system this is very small, $\delta \tilde r \approx 4.4$~m, but it gets larger for the outer planets, for example for the Sun-Jupiter system we have $\delta \tilde r\approx11$~Km. The Earth-Moon system gives $\delta \tilde r\approx1.6$~m, but this is actually the best ratio $\delta \tilde r/R$ of the three examples.

There is a much larger intermediate region where  there are significant perturbative corrections to Newtonian theory, but where deep MOND behavior is not yet in evidence. However this transition region is very model dependent.  For the model introduced in~\cite{teves} Milgrom's $\tilde\mu$ can be estimated in the quasi-Newtonian region through formula (69) there (which formula, however, is rigorous only in the spherically symmetric case): \be              \label{mubar} {\tilde \mu}={F^{(N)}\over F} \approx{\left(1-{16\pi^3\over k^3}{a_0^2\over F^2} \right)}. \ee               Let us take $F\approx F^{(N)}$ and use Eq.~(\ref{newtonian}).  We  see that $10^{-4}$ departures from Newtonian gravity occur within a semi-major axis of size
\be              \label{deltar-4}
\Delta \tilde r={800\pi^{3/2}\over k^{3/2}}{a_0\over A}\, . \ee             Using $k\approx 0.03$ (as suggested in~\cite{teves}) this is $\Delta \tilde r\approx 1,900$~Km, $\Delta \tilde r\approx 4.7\times 10^6$ Km, and $\Delta \tilde r\approx 700$~Km, respectively, for the three examples described above.

Of course the calculation in this section is very idealized.  Orbits are not circular, the center of mass does not coincide with the Sun center, and rather than two objects we have a multitude of competing influences. In Section~\ref{total} we consider more realistic situations, examining also the surprising effect of the extra solar component (negligible in Newtonian theory but not in MOND). Before that, however, we must reconsider the effect of the neglected curl term. We do this in Sections~\ref{curlsec}-\ref{extreme}. While the qualitative aspects of the present section survive the introduction of the curl term, we shall find that the quantitative conclusions are significantly modified.

\section{Accounting for the curl term}\label{curlsec}

By carrying out the curl in Eq.~ (\ref{second}) we get\be              \nabla\ln \mu\wedge\mathbf{u}-\nabla\wedge\mathbf{u}=0,\label{curl}\ee              while squaring Eq.~(\ref{u}) gives\be              u^2=(4\pi/k)^2\mu^2|\nabla\phi|^2.\label{u2}\ee              In T$e$V$e$S $\mu=\mu(k|\nabla\phi|/a_0)$; thus $k^4 u^2/a_0{}^2$ is a function of $\mu$ only.  Defining the dimensionless quantity\be              \kappa\equiv \partial \ln u^2/\partial\ln\mu,\ee              we get $\nabla\ln\mu=\kappa^{-1}\nabla u^2/u^2$ so that Eq.~(\ref{curl}) becomes\be              \kappa\,u^2\nabla\wedge \mathbf{ u}+\mathbf{ u}\wedge\nabla u^2=0.\label{basic}\ee

In systems with spherical, cylindrical or planar symmetry,
$\mathbf{u}$ is necessarily collinear with $\nabla|{\mathbf
u}|^2$.  Then $\nabla\wedge \mathbf{ u}$ must vanish everywhere
(since $\kappa$ would be expected to vanish only at isolated
points). This agrees with the findings of
Refs.~\cite{aqual}-\cite{teves} that $\mu\nabla\phi$ and
$\mu\nabla\Phi$ are both curl free in such situations.   When the
spatial symmetry is lower or nonexistent, the second term in
Eq.~(\ref{basic}) will not generally vanish, and will be of order
$|\mathbf{u}|^3/L$ where $L$ denotes the scale on which
quantities vary.  Thus if in a region $\kappa\gg 1$, we would
expect $|\nabla\wedge \mathbf{u}|$ to be much smaller than its
expected magnitude $|\mathbf{u}|/L$; this signals the
quasi-Newtonian regime where $\mathbf{u}$ is nearly curl-free.

\begin{figure}[h]
\centering
\includegraphics[width=6cm]{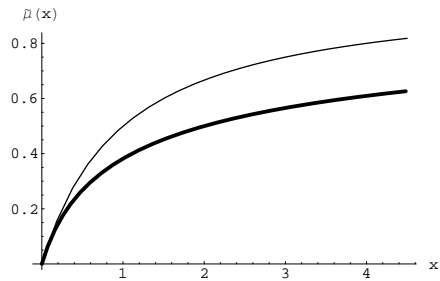}
  \caption{Milgrom's $\tilde\mu$ function (upper curve) as used in many extragalactic applications (Eq.~(\ref{tildemu1}))  and as defined in this paper with $k=0.03$ by Eqs.~(\ref{tildemu2})-(\ref{x}). }
 \label{figure:fig1}
 \end{figure}

In T$e$V$e$S the manner of transition between the deep MOND and
Newtonian regimes  is  dependent upon the form of $\mu$.  The form
proposed in Ref.~\cite{teves} is quite difficult to work with in
our context.  We shall thus replace it by the implicit expression
\be              {\mu\over \sqrt{1-\mu^4}}= {k\over 4\pi}{|\nabla\phi|\over a_0}. \label{muhere} \ee               which satisfies the limit (\ref{mumond}).   A simple calculation (see Ref.~\cite{teves}) then shows that Milgrom's $\tilde\mu$, defined by Eq.~(\ref{AQUAL}), is here given parametrically by \bea \tilde\mu&=&\zeta(1+\zeta)^{-1}\label{tildemu2}\\x&=&\zeta(1+\zeta)\left[1-(k\zeta/4\pi)^4\right]^{-1/2}\label{x}\eea which satisfies the MOND requirements that $\tilde \mu\to x$ for $x\ll 1$ and $\tilde \mu\to 1$ as $x\to \infty$.  Fig.~\ref{figure:fig1} compares our $\tilde\mu$ with the ``simple'' $\tilde\mu$ of Refs.~\onlinecite{milg1} and \onlinecite{zhaofam}.

Eliminating $|\nabla\phi|/a_0$ between Eqs.~(\ref{u2}) and (\ref{muhere}) gives
\be              {u^2\over a_0{}^2}={256\pi^4\over k^4}{\mu^4\over 1-\mu^4}\,.
\label{usquared}
\ee
Differentiating the logarithm of this we calculate that
\be              \kappa={4\over 1-\mu^4}=4+{k^4\over 64\pi^4}{u^2\over a_0{}^2}\,.
\label{kappa}
\ee

In terms of the dimensionless vector field
\be
\mathbf{U}\equiv {k^2\over 16\pi^2}{\mathbf{u}\over a_0}
\label{U}
\ee
we may thus cast Eqs.~(\ref{basic}) and (\ref{first}) into the form
\bea
\nabla\cdot \mathbf{U}=0
\label{a}
\\
4(1+U^2)\,U^2\,\nabla\wedge \mathbf{ U}+\mathbf{ U}\wedge\nabla
U^2=0. \label{b} \eea
where we have dropped the source of the
first since we  are interested only in the region near the SP.
This pair of exact equations for one dimensionless vector is
central to our study.

Once $\mathbf{U}$ is solved for we can recover $\nabla\phi$ by
combining Eqs.~(\ref{u}), (\ref{usquared}) and (\ref{U}):
\be              -\nabla\phi={4\pi a_0\over k}(1+U^2)^{1/4}{\mathbf{U}\over
U^{1/2}}\,. \label{gradphi} \ee
As remarked earlier, the condition $\kappa\gg 1$ brings in the Newtonian
limit.  Now $\kappa\gg 1$ is equivalent to $U\gg 1$.  Obviously in
this case  $-\nabla\phi\approx (4\pi a_0/k){\mathbf
U}=(k/4\pi)\mathbf{u}$ which tells us by Eq.~(\ref{u}) that
$\mu\approx 1$, indeed the Newtonian limit (the same is obvious
from Eq.~(\ref{kappa})).

\section{The quasi-Newtonian region}\label{quasi}

At this point we go over to spherical polar coordinates
$(r,\psi,\phi)$ with  origin at the SP; accordingly
\be               z= r
\cos\psi; \qquad \varrho= r\, \sin\psi.
\ee               So, for example,
Eq.~(\ref{newtonian}) takes the form
\be               \label{newtonian1} {\mathbf F}^{( N)}=Ar{\mathbf N}, \ee
where
\bea {\mathbf N}(\psi)&\equiv & N_r {\mathbf  e}_r
+N_\psi{\mathbf e}_\psi \label{newt1}
\\
N_r&=&{\scriptstyle 1\over\scriptstyle 4}[1+3\cos(2\psi)]\\
N_\psi&=&-{\scriptstyle 3\over\scriptstyle 4}\sin(2\psi).
\label{newt2}
\eea

We define the quasi-Newtonian region as that where  $U^2$ is of
order 1 or  larger so that the factor $1+U^2$  cannot be ignored
in  Eq.~(\ref{b}). The region's size  may be estimated by dropping
the curl term in (\ref{ufirst}) (an approximation to be justified
a posteriori) and finding the solution to $U^2=1$ using
(\ref{newtonian}) and ({\ref{U}) (in the Newtonian region
${\mathbf u}={\mathbf F}^{(N)}$). This leads to the ellipsoid:
\be              \label{r0ellipse} r^2\Big(\cos^2\psi+{1\over 4}\sin^2\psi
\Big)=r_0^2\equiv \Big({16 \pi^2 a_0\over k^2 A}\Big)^2 \ee
Eq.~(\ref{b}) tells us that
well outside of this ellipsoid the curl is suppressed by a factor of $1/r^2$
with respect to ${\mathbf F}^{(N)}$.  As we show below
${\mathbf U}$ is then neatly separated into a Newtonian component ${\mathbf U}_0$  (carrying the divergence predicted by (\ref{first}) and depicted in Fig.~\ref{Nfield})
and a ``magnetic'' component ${\mathbf U}_2$. By definition ${\mathbf U}_2$ is solenoidal and to
leading order is sourced purely by ${\mathbf U}_0$.  Specifically
the dynamics is approximated by
\bea
{\mathbf U}&=&{\mathbf U}_0+{\mathbf U}_2\\
{\mathbf U}_0&=&{r\over r_0}{\mathbf N}(\psi)
\label{U0r0}\\
\nabla\cdot {\mathbf U}_2&=&0\label{div2}\\
\nabla\wedge {\mathbf U}_2&=&-{{\mathbf U}_0\wedge\nabla
|{\mathbf U}_0|^2\over 4
|{\mathbf U}_0|^4}
\label{curl2}
\eea
\begin{figure}
\centering
\includegraphics[width=6cm]{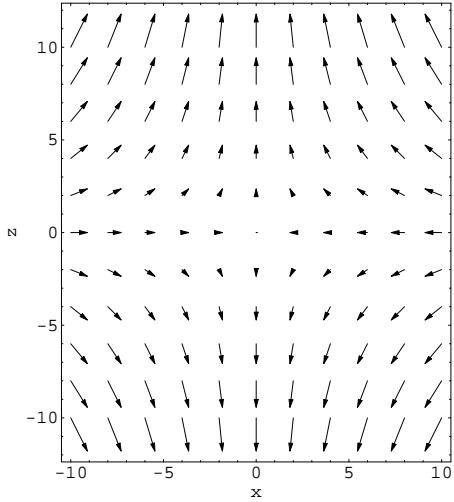}
  \caption{The flow of ${\mathbf U}_0$ around the SP (at the origin) in a plane containing the symmetry ($z$) axis; for clarity all vectors have been linearly rescaled. }
\label{Nfield}
\end{figure}
With the notation
\be
{\mathbf U}_2=U_r\ve_r+U_\psi\ve_\psi
\label{U2}
\ee
Eqs.~(\ref{div2}) and (\ref{curl2}) become
\bea
{1\over r^2}{\partial\over\partial r}(r^2 U_r)+{1\over r\sin\psi}{\partial\over
\partial\psi}(\sin\psi U_\psi)&=&0,
\label{div3}\\
{1\over r}\left[{\partial\over\partial r}(r U_\psi)-{\partial U_r\over \partial \psi}\right]&=& {s(\psi)\over r^2},
\label{curl3}
\eea
with
\be
s(\psi)\equiv -
{3\over 8}{\cos\psi\sin\psi\over {\left[\cos^2\psi+{\sin^2\psi\over 4}
\right]}^2}=-{12\sin 2\psi\over (5+3\cos2\psi)^2}\,.
\ee
The form of Eqs.~(\ref{div3}) and (\ref{curl3}) suggests that both $U_r$ and $U_\psi$ behave as $1/r$.  Accordingly we recast Eq.~(\ref{U2}) as the ansatz
\be
\label{U2anz}
{\mathbf U}_2={r_0\over r}{\mathbf B}(\psi)={r_0\over r}\Big({F(\psi)\ve_r+G(\psi)\ve_\psi}\Big),
\ee
where the $r$ dependence has been fully factored out.
With this ansatz Eq.~(\ref{curl3}) collapses into
\be
F'=-s={12\sin 2\psi\over (5+3\cos2\psi)^2}\,,
\ee
with solution
\be
F={2\over 5+3\cos2\psi}+A,
\label{sol1}
\ee
where $A$ is a constant. Eq.~(\ref{div3}) now becomes
\be
F+{1\over \sin\psi}{\partial\over
\partial\psi}(\sin\psi\,  G)=0,
\ee which integrates to  \be G\sin\psi=-\int F\sin\psi\, d\psi +B
\ee where $B$ is another constant. Performing the integral gives
\bea G\sin\psi&=&{\tan^{-1}(\sqrt 3 -2 \tan{\psi\over 2})+
\tan^{-1}(\sqrt 3 +2 \tan{\psi\over 2})\over \sqrt 3}\nonumber\\
&& + A\cos\psi+B.
\eea

\begin{figure}
\centering
\includegraphics[width=6cm]{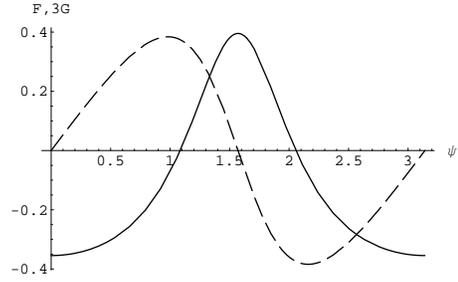}

 \caption{The angular profile functions $F$ (solid) and $G$ (dashed) giving the direction of the ``magnetic'' field ${\mathbf B}$
in the quasi-Newtonian region; for clarity $G$ has been multiplied by 3.}

\label{FGfarout}

\end{figure}

To determine $A$ and $B$ we must discuss boundary conditions.  According to Milgrom~\cite{milg2} for the system (\ref{a})-(\ref{b}) the normal component of ${\mathbf u}$ (or ${\mathbf U}$) must vanish on all boundaries.  Parts of the symmetry axis ($\psi=0$ as well as $\psi=\pi$) are evidently a boundary of the quasi-Newtonian region; it is obvious that $N_\psi$ vanishes on both North and South parts of it, where it is the normal component.  Thus since ${\mathbf U}_0$ satisfies the boundary condition on the relevant pieces of the axis, so must ${\mathbf U}_2$.   Accordingly  we must require $G(\psi=0)=G(\psi=\pi)=0$, from which follows that
\be
A=B=-{\pi\over 3\sqrt 3}\,.
\ee
The solutions $F$ and $G$ are plotted in Fig.~\ref{FGfarout}.
We find that $G(\pi/2)=0$ as well.  Thus on the  symmetry plane ($\psi=\pi/2$) ${\mathbf U}$ is collinear with the axis.

\begin{figure}

\centering

\includegraphics[width=6cm]{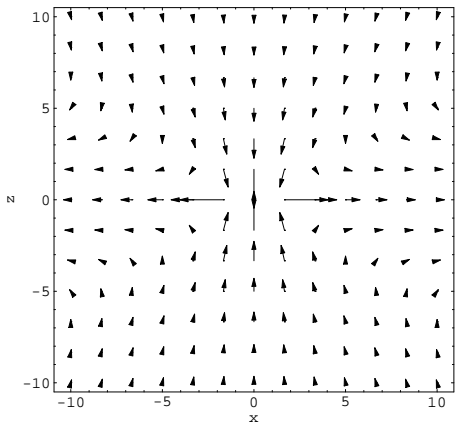}

  \caption{The flow of ${\mathbf U}_2$ in a plane containing the $z$ axis; coordinates are  in units of $r_0$.   For clarity the solution was cut off at $r=r_0$ (so as to avoid a
divergence at the origin).}

\label{Bfield}

\end{figure}

What about the rest of the boundary?  We see from
Eq.~(\ref{U2anz}) that ${\mathbf U}_2\rightarrow 0$ as
$r\rightarrow \infty$.  Thus at large $r$ our ${\mathbf U}$ merges
with ${\mathbf U}_0$ which we know to be the limiting form of the
Newtonian field as we approach the SP.  It follows that our
solution automatically fulfills the boundary conditions at large
$r$.  The inward part of the boundary of the quasi-Newtonian
region adjoins the intermediate MOND region, where MOND effects
are no longer small.  Fortunately there is no need for us to set
boundary conditions there; rather, the solution just described
serves to set boundary conditions for the intermediate MOND
region.

We conclude that a SP far away from the strong MOND bubble is
characterized by a Newtonian component proportional to $r$
together with a magnetic-like perturbation that falls off like
$1/r$. The full physical effects in this  regime may be
appreciated by combining (\ref{gradphi}) with (\ref{U0r0}) and
(\ref{U2anz}). We find that the extra acceleration felt by test
particles is
\be      \delta{\mathbf F}=-\nabla\phi\approx {4\pi
a_0\over k}{\left({\mathbf U_0} +{{\mathbf U_0}\over
4U_0^2}+{\mathbf U_2}+\cdots \right)}. \ee
The first contribution, call it $\delta{\mathbf F}_0$ is of fully Newtonian form, and just serves to renormalize the gravitational
constant, as discussed in~\cite{teves}. The second term was also
derived in~\cite{teves} (c.f. Eq.~(69) of \cite{teves}) and is
\be               \label{deltaF1}\delta{\mathbf F}_1={16\pi^3\over
k^3}{a_0^2\over F^{(N)}{}^2}{\mathbf F}^{(N)} ={8\pi a_0\over
k}{r_0\over
r}{{\mathbf N}(\psi)\over 5+ 3\cos(2\psi)}. \ee
What we have just
shown is that to these two terms one should add the magnetic-like
contribution
\be              \label{deltaF2} \delta{\mathbf F}_2 ={4\pi a_0\over k}{r_0\over
r}{\mathbf B}(\psi), \ee
which is of the same order of magnitude as
$\delta{\mathbf F}_1$.   Apart from the prefactor $4\pi a_0/k$, this term is just what was plotted in Fig.~\ref{Bfield}.
In Fig.~\ref{corrections} we plot the angular profile  ${\mathbf
B}(\psi)+2[5+3\,\cos(2\psi)]^{-1}{\mathbf N}(\psi)$ of the total
correction to the acceleration after renormalization of $G$.  The plotted field  is to be
divided by $r$ (and multiplied by $4\pi a_0 r_0/k$) to obtain the
extra acceleration felt by test particles in the quasi-Newtonian
region.

How do these results affect the naive expectations of
Section~\ref{bubbles}? We have just shown that a full quantitative
analysis can never neglect the ``magnetic'' field derived in this
Section. In addition the border between full  and linear MONDian
behavior is determined by the condition $U^2=1$, equivalent to
ellipsoid (\ref{r0ellipse}). As long as we stay well outside this
ellipsoid we obtain results consistent with (\ref{mubar}) and
(\ref{deltar-4}); however the order of magnitude of linear
corrections outside this ellipsoid may be written
\be              \label{linpercent}
 {\delta F\over F^{(N)}}\sim {\left(4\pi\over
k\right)}^3{\left(a_0\over A\right)}^2{1\over r^2}= {k\over
4\pi}{\left(r_0\over r\right)}^2 \, . \ee
We learn that the highest fractional correction in this regime is
achieved close to ellipsoid (\ref{r0ellipse}) and is of order
$k/(4\pi)$, around a $0.0025$ for $k\approx 0.03$; it
then falls off as $1/r^2$ as we move away from the SP.
Therefore, as long as we do not use (\ref{mubar}) for fractional
corrections larger than $k/(4\pi)$ we obtain qualitatively correct results (the example given in Section~\ref{bubbles} satisfies
this condition).

The bottom line for our predictions is that the ellipsoid
(\ref{r0ellipse}) represents both the region where the largest
linear corrections are felt and the border for the onset of full
MOND behavior. For the three examples considered in
Section~\ref{bubbles} we have \bea
r_0&\approx &383\  {\rm Km\qquad Earth-Sun}\\
 r_0&\approx &9.65\times 10^5\ {\rm Km\qquad Jupiter-Sun}\\
r_0&\approx &140\ {\rm Km\qquad Earth-Moon} \eea      corresponding to
ellipsoids with major semi-axis of  766 Km (Sun-Earth), 1.93$\times 10^6$
Km (Sun-Jupiter) or 280 Km (Earth-Moon). These are
the relevant dimensions of the MOND bubbles.

\begin{figure}
\centering
\includegraphics[width=6cm]{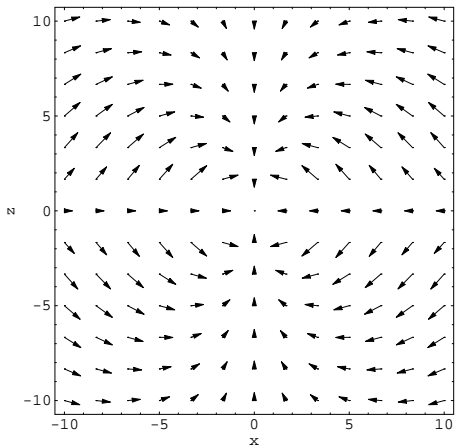}
\caption{The flow of  $\mathbf{B}(\psi)+2[5+3\,\cos(2\psi)]^{-1}{\mathbf N}(\psi)$. When divided
by $r$ and multiplied by $4\pi a_0 r_0/k$ this field gives the
physical acceleration beyond the Newtonian one felt by test
particles in the quasi-Newtonian region.} \label{corrections}
\end{figure}

 \section{The deep MOND region}\label{extreme}

By Eq.~(\ref{kappa}) the deep MOND regime ($\mu\ll 1$) entails $\kappa\approx 4$ or $U\ll 1$.  Thus  in Eq.~(\ref{b}) we
replace $1+U^2\to 1$.  Then together with Eq.~(\ref{a}) this has a double symmetry, already noticed by Milgrom~\cite{milg3}.  They are both invariant under $\mathbf{U}\to {\rm const.}\times \mathbf{U}$ (rescaling), and under ${\bf x}\to \lambda\mathbf{x}$ (dilation of the coordinates).  The first symmetry implies that the normalization of $\mathbf{U}$ is arbitrary (of course the normalization is eventually fixed by taking cognizance of the sources of Eq.~(\ref{a})).  The second means that a solution whose linear scale is expanded remains a solution.

In spherical polar coordinates Eqs.~(\ref{a})-(\ref{b}) take the form
\begin{eqnarray}
{1\over r^2}{\partial\over\partial r}(r^2 U_r)+{1\over r \sin\psi}{\partial\over\partial\psi}(\sin\psi\, U_\psi)=0
\label{U*}
\\
\left[{4\over r}\Big( {\partial (r U_r)\over\partial r}-{\partial U_\psi\over \partial\psi} \Big)+\Big({U_r \over r}{\partial\over \partial\psi}-U_\psi{\partial\over\partial r}\Big)\right]U^2=0
\label{V*}
\end{eqnarray}

For a solution of these to turn into a second solution upon dilatation of the coordinates ($r\rightarrow \lambda\, r$), it is necessary for the $r$ dependence of both $U_r$ and $U_\psi$ to be a single power.  Thus we make the ansatz
\begin{equation}
 \mathbf{U}=C\Big({r\over r_0}\Big)^{\alpha-2}(F(\psi)\,
\mathbf{ e}_r+G(\psi)\,\mathbf{ e}_\psi)
\label{Ufinal}
\end{equation}
with $C$ and $\alpha$ dimensionless constants. The power
$\alpha-2$ was chosen for notational convenience in what follows.
Substituting in Eq.~(\ref{U*}) we obtain
\be G'+{\rm ctan}(\psi)\, G+ \alpha F=0 \label{f} \ee
while substitution in Eq.~(\ref{V*}) gives
\be
 F{d(F^2+G^2)\over d\psi}+ 2\big[\alpha G-2F'\big] (F^2+G^2)=0.
 \label{g}
\ee These last  constitute a coupled system of first order
ordinary differential  equations for $F(\psi)$ and $G(\psi)$.

These equations have several symmetries. $F$ and
$G$ may be rescaled, that is multiplied by a constant (this is nothing but the scale-invariance of the deep MOND regime). We also
have the symmetry: $ \alpha \rightarrow -\alpha$, $F\rightarrow F$,
$G \rightarrow -G$. Finally the equations are parity-invariant:
\bea
\psi&\rightarrow& \pi - \psi,
\nonumber\\
F&\rightarrow& \pm F,
\nonumber\\
G&\rightarrow& \mp G\, .
\eea
Of course this by itself does not compel the solutions themselves to have definite parity, i.e. $F(\psi)=\pm F(\pi-\psi)$ and $G(\psi)=\mp
G(\pi-\psi)$. However, numerically we  find that the only regular solutions are indeed those with definite parity, and that these only exist for a discrete sequence of $\alpha$s: $\{\pm \alpha_1, \pm \alpha_2,\cdots\ \}$. Specifically we find
$\alpha_1=2$ and the approximate values $\alpha_2\approx 3.528$,
$\alpha_3\approx 5.039$, $\alpha_4\approx 6.545$, etc.

\begin{figure}\centering
\includegraphics[width=6cm]{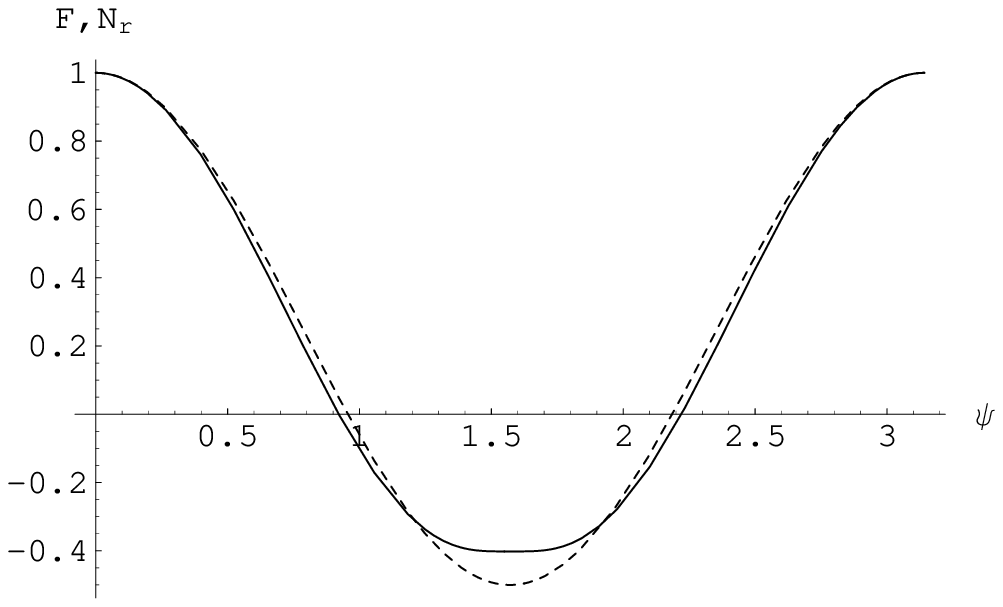}
\includegraphics[width=6cm]{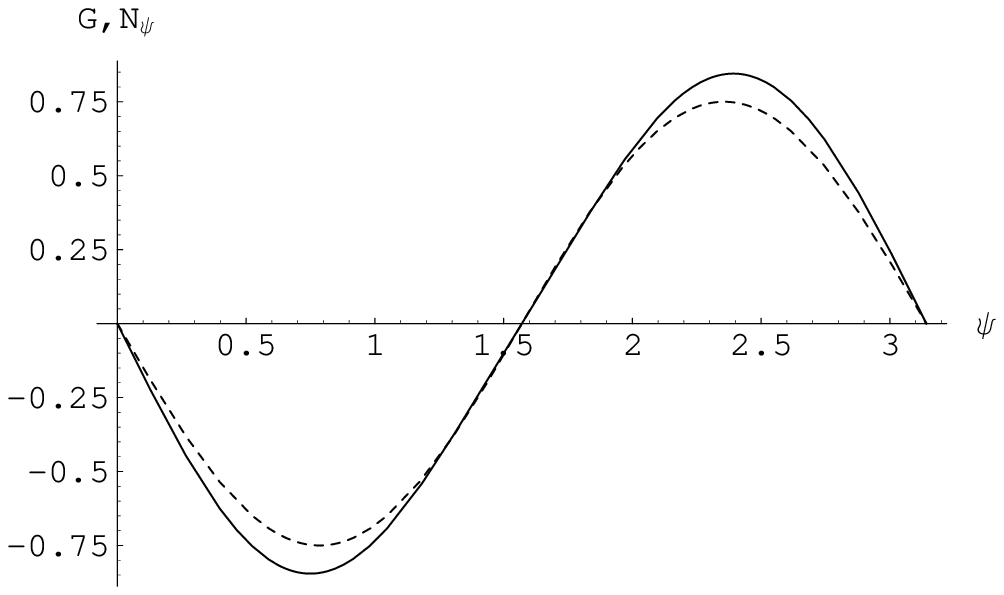}
\caption{The numerically determined angular profile functions $F$
and $G$ in the deep MOND region (solid) compared with the
Newtonian profile functions $N_r$ and $N_\psi$ (dotted),
respectively.} \label{FGmond}
\end{figure}

Seen in another
way, the boundary conditions at $\psi=0$ (see below) justify representing $F$ as a Fourier series in $\cos(m\psi)$ and $G$ as a Fourier series in $\sin(m\psi)$. It is only for the mentioned special $\alpha$s  that even and odd $m$ modes decouple, so that we can have a solution that is a series in only odd or only even $m$. For
other values of $\alpha$ the solutions mix even and odd $m$, but are
singular at $\psi=\pi$.

We further found numerically that each of the regular solutions is
dominated by a single Fourier mode, i.e. it is of the form
$F\approx F_0+F_n \cos n\psi$ and $G\approx G_n\sin n\psi $ for a
given $n$. The solution corresponding to $n=1$ is obtained for
$|\alpha_1|= 2$ and is exactly $F=a\cos\psi$ and $G=\mp
a\sin\psi$ ($a$ is a constant) for $\alpha=\pm 2$, respectively. We have been unable to find analytic expressions for other coefficients $F_n$ and $G_n$, but have determined them
numerically. We now select the relevant solution
by  imposing appropriate boundary conditions for our problem.

As in Sec.~\ref{quasi} the  boundary condition that the normal component of  $\mathbf{U}$ vanish requires  that we take $G(\psi=0)=G(\psi=\pi)=0$.  Because $C$  can  still be adjusted, we loose no generality in requiring the corresponding
boundary condition $F(\psi=0)=F(\psi=\pi)=1$. For were we to demand
$F(\psi=0)\neq F(\psi=\pi)$, we would thereby introduce a
jump in $\mathbf{U}$ across the plane $\psi=\pi/2$ for which
there is no physical reason.  Our choice of boundary conditions immediately selects a solution with definite
parity, which as mentioned earlier, are the only nonsingular ones. Regarding boundary conditions at large $r$,
we know that there must be a match with the field in the
quasi-Newtonian region. This naturally selects the particular solution with $n=2$, since the quasi-Newtonian solution $\mathbf{U}_0$ has components with angular profiles of form $\cos 2\psi$ or $\sin 2\psi$. This logic still does not prefer positive over negative $\alpha$.  But to avoid a singularity at the origin (see Eq.~(\ref{Ufinal})) we should select the solution with positive $\alpha$, namely that for $\alpha\approx 3.528$.

The functions $F$ and $G$ obtained for this
$\alpha$ are plotted in Fig.~\ref{FGmond}. These graphs are
approximated at the level of 1 \% by the formulae
\bea \nonumber
F(\psi)&=&0.2442+ 0.7246 \cos(2\psi) + 0.0472 \cos(4\psi),
\\
G(\psi)&=&-0.8334 \sin(2\psi) - 0.0368 \sin(4\psi).
\eea

For  comparison Fig.~\ref{FGmond} plots also $N_r$ and $N_\psi$ of
Eqs.~(\ref{newt1})-(\ref{newt2}).  We see that the angular profile
of the deep MOND $\mathbf{U}$ (whose  flow is plotted in
Fig.~\ref{Umond}) is quite similar to that of the Newtonian
$\mathbf{U}_0$ (Eq.~(\ref{U0r0}) and Fig.~\ref{Nfield}).  Of
course, the radial dependences of the two are quite different.
Now as mentioned earlier, in the absence of any mention of the
sources in Eqs.~(\ref{a})-(\ref{b}), it is not possible to
determine the normalization of $\mathbf{U}$.  However, we may
estimate $C$ in Eq.~(\ref{Ufinal}) as follows.  Given the
similarity of the angular profiles we may suppose that were we to
extend the deep MOND $\mathbf{U}$ of Eq.~(\ref{Ufinal}) to the
inner boundary of the  Newtonian region at $r=r_0$, we should
obtain $\mathbf{U}_0$.  This requires that $C=1$ and we adopt
this value.

\begin{figure}
\centering
\includegraphics[width=6cm]{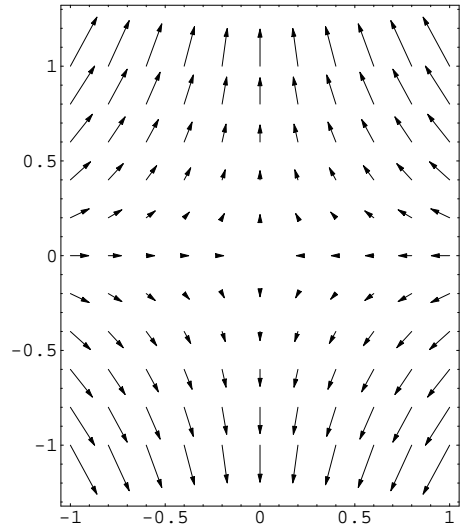}
\caption{The flow of the field $\mathbf{U}$ in the deep MOND regime plotted with linear scale in units of $r_0$ and assuming $C=1$.}
\label{Umond}
\end{figure}

We conclude that taking the curl term into account in the deep
MOND regime once again vindicates qualitatively the simplified
arguments of Section \ref{bubbles}, but introduces substantial
quantitative novelties. Using (\ref{gradphi}) we find that the
extra physical force is now
\be \delta \mathbf{F}\approx -\nabla \phi={4\pi a_0\over
k}{\mathbf{U}\over U^{1/2}}\,. \ee
If we define $\mathbf{D}$ as the angular profile in the deep MOND
regime (which as we have seen is very close to $\mathbf{N}$) then
\be\label{delfdeep} \delta \mathbf{F}\approx {4\pi a_0\over
k}{\left(r\over r_0\right)}^{\alpha-2\over 2}{\mathbf{D}\over
D^{1/2}}\,. \ee
For $\alpha<4$ (a condition satisfied by our solution), the tidal
stresses associated with this field, i.e its spatial derivatives,
diverge at the saddle, as predicted in Section~\ref{bubbles}.
However the divergence is softer than in Eq.~(\ref{approx}) where
the curl term was ignored (that solution corresponds to
$\alpha=3$, which is an unphysical value as we have seen).

Rearranging (\ref{delfdeep}) with the aid of the definition of
$r_0$ and Eq.~\ref{newtonian1} we find that the continuation of
formula (\ref{linpercent}) in the deep MOND regime is
\be
 {\delta F\over F^{(N)}}\sim {k\over
4\pi}{\left(r_0\over r\right)}^{\alpha -4 \over 2} \approx {k\over
4\pi}{\left(r_0\over r\right)}^{-0.24}\, .
\ee
Hence the fractional correction to Newtonian gravity, which equals
$k/(4\pi)\approx 0.0025$ at the ellipsoid (\ref{r0ellipse}),
continues to grow in the strong MOND regime as we approach the
saddle. (Were we to ignore the curl term, in which case $\delta F/F^{(N)}\sim 1/\sqrt{r}$, this growth would be steeper.) One
implication of the growth is that the $\phi$ force overtakes $F^{(N)}$ in a much smaller inner region than  naively expected (cf. formulas (\ref{deltara0})).  Specifically $\delta F\approx F^{(N)}$ at
\be
r\sim r_0 {\left(k\over 4\pi\right)}^{2\over 4-\alpha }={a_0\over
A}{\left(k\over 4\pi\right)}^{2{\alpha-3 \over 4-\alpha} }.
\ee
This is smaller than (\ref{deltara0}) by a factor of $10^{-6}$,
and is essentially microscopic except for the Jupiter-Sun system.
The value of $F^{(N)}$ when it becomes subdominant is not $a_0$ as
naively expected; it is also smaller by a factor of $10^{-6}$.

In summary,  the full analysis reveals that there is a very large
region (given by the ellipsoid (\ref{r0ellipse})) inside which
full MONDian effects are present. The fractional MONDian corrections to gravity in this region exceed $k/(4\pi)$ and are therefore
significant. However the MOND field only dominates the Newtonian
field, i.e. the fractional correction becomes larger than unity, in a
region far too small to be observable.

\section{The realistic solar system}\label{total}

The results of Secs.~\ref{bubbles}, \ref{quasi} and \ref{extreme} can be used to show that the SP location for a pair of masses as determined by pure Newtonian gravity ($\mathbf{F}^{(N)}=-\nabla\Phi_N=0$) coincides with that determined by full T$e$V$e$S ($-\nabla(\Phi_N+\phi)=0$).  In the calculations in Sec.~\ref{extreme} the origin $r=0$ is the point where $\nabla\phi=0$, and the field configuration of $\nabla\phi$, or its surrogate $\mathbf{U}$, in North and South hemispheres are reflections of each other (see Fig.~\ref{Umond}).  This configuration acts as a boundary condition for $\mathbf{U}$ in the quasi-Newtonian region (treated in Sec.~\ref{quasi}).  Accordingly, we expect not only  the ``magnetic'' part $\mathbf{U}_2$, but also the $\mathbf{U}_0$, which serves as background for $\mathbf{U}_2$'s equation (\ref{curl2}), to reflect the mentioned symmetry, and for the null points of both these fields to coincide with that of $\mathbf{U}$ of the deep MOND region.  Now as we move outward from the quasi-Newtonian region, $\mathbf{U}$ becomes dominated by $\mathbf{U}_0$ which is the pure Newtonian field.  Hence the SP determined by that field (see Sec.~\ref{bubbles}) coincides with that determined by the full MOND field.

The results presented so far are ``model calculations'', valid under a number of simplifying assumptions not satisfied by the real SS.
For example, orbits are elliptic, not circular; the barycenter of the system does not coincide with the center of $M$; we have a many-body problem, not a two-body problem; etc.   To leading order these complications do not change the
anomalous effects predicted by MOND around SPs or the size of
the regions where they are felt.
They do complicate the issue of locating ``MOND bubbles'', but since
their centers coincide with the SPs of the Newtonian potential,
this is in fact a Newtonian physics problem, independent of MOND dynamics.

For example the SS barycenter  is dominated by the
Sun-Jupiter pair and lives just outside the solar surface, rotating with a
period of approximately 11 years. But even this is a crude approximation:
the relative position of the Sun and any planet depends on
the configuration of the entire SS, and is chaotic.  The same may be said for the location of the SP between the Sun and that planet.
However, with empirical inputs and a numerical Newtonian code we can determine
the location of the full set of SPs, and even predict
where they will be within a few years~\cite{boulder}.
Not only are these details in the realm of Newtonian physics,
but they do not affect our conclusions on
MOND effects around SPs, as long as we anchor our solutions to
wherever Newtonian theory predicts the SPs to be. Indeed
we only need the result (\ref{newtonian}) as
a boundary condition for our MONDian calculations.

\subsection{An example: Epicycles and the many-body problem}

However, these practical details do affect the planning of experiments,
as we now illustrate. Consider the location of the Earth-Sun and
Moon-Earth SPs. Ignoring the effect of the Moon, the Earth-Sun
SP is predicted to be  well inside the Moon's orbit, so we cannot
decouple the two systems. This induces dramatic qualitative
changes in the location of both SPs.

Let us assume that within the Moon-Earth orbit the Sun's field
is uniform (with strength $a_E$)
and parallel to the Lunar orbital plane (this may be refined,
but does not alter our point). If we consider a frame
attached to the unperturbed Moon-Earth SP, with $z$ pointing away from
the Earth and $x$ on the orbital plane, then the Newtonian acceleration
for points on this plane is
\be\label{FN}
\mathbf{F}_N=A\big(z\mathbf{e}_z-{\scriptstyle 1\over \scriptstyle 2}x\mathbf{e}_x\big)
-a_E[\cos (\omega t)\mathbf{e}_z-\sin(\omega t)\mathbf{e}_x],
\ee
where $A$ is given by Eq.~(\ref{A}) with $m$ the Moon's mass and $M$ the Earth's, $\omega=2\pi/T$, $T$  the Moon's synodic period, and $t=0$ the time when the Sun, Earth and Moon are aligned (in that sequence). Thus, within these
approximations, the actual SP (defined by $\mathbf{F}_N=0$)
describes a monthly ellipse centered on the unperturbed SP location,
with equation
\bea
x&=&2{a_E\over A}\sin(\omega t),
\\
z&=&{a_E\over A}\cos(\omega t).
\eea
Thus the semimajor axis is $2 a_E/A\approx 6.0\times 10^4\,$Km while the
semiminor one is $a_E/A\approx 3.0\times 10^4\,$Km;
these axes perform an annual rotation so that the smaller axis
stays aligned with the Sun. Since the rough prediction
(ignoring the Sun) places the SP
about $4.3\times 10^4\,$Km away from the Moon, we cannot ignore
the details arising from the three-body problem. When all effects are taken into account, the SP may encounter the Moon's surface in its motion.

Other perturbations superpose further ellipses upon this motion,
each aligned with the direction of its source. In general the motion
of the SP is a series of elliptic epicycles of this form
due to all possible perturbations on the main two-body system.

It should be pointed out that  the fact that the Earth-Moon system falls freely in the field of the Sun does not alter the above arguments (see Sec.~\ref{capsule}).  In spite of the weak equivalence principle, the criterion for strong MOND effects is that the field $\mathbf{F}^{(N)}$ calculated in the global frame becomes comparable to $a_0$ so that $\nabla\phi$ can make a significant and identifiable contribution to the overall gravitational field $-\nabla(\Phi_N+\phi)$.  Hence the position of the SP is to be determined, to first order,  as done above.

\subsection{MOND bubbles as accelerometers}\label{extra}

There is a fine detail  of the SP system
that is purely MONDian: sensitivity
to the extra-solar potential, or more precisely, to the peculiar
acceleration, $a_p$, of the SS barycenter.
As explained before, MONDian behavior can only be identified
from the total potential and this {\it must} include the extra-SS component.
In Newtonian theory the effect of $a_p$ passes unnoticed because nothing
dramatic distinguishes the SP; by contrast in MOND the SP is
signaled by diverging tidal forces. The fact that we are
free-falling in this  field is irrelevant as just mentioned.  We note that $a_p$ does not shift the location of the
Lagrange points because their definition involves balancing inertial
forces and gravitational fields. Since we are free-falling in the extra-galactic field, aside from making a tidal distortion, this acceleration does not affect the Lagrange points.

The effect of $a_p$ is to superpose a further elliptical motion onto the
larger epicycles due to intra-solar perturbations.
If $\mathbf{a}_p=\{a_{pz},a_{px},a_{py}\}$  in the system of axes
used above, then a similar calculation leads to
\bea
z&=&{2\over A}(a_{pz}\cos(\omega' t)-a_{px}\sin(\omega' t)),\\
x&=&{2\over A}(a_{pz}\sin(\omega' t)+a_{px}\cos(\omega' t)),\\
y&=&{2\over A}a_{py},
\eea
where $\omega'$ is $2\pi$ divided by the Moon's \emph{sidereal} period.  We see that the SP is raised off the orbital plane by $a_{py}/ A$,
and describes an ellipse oriented with $\mathbf{a}_p$ on this plane.

How can we estimate $\mathbf{a}_p$? This is the ``acceleration
counterpart'' to the CMB dipole (which measures the peculiar velocity
$\mathbf{v}_p$ with respect to the cosmological frame). There is no simple
way to estimate $\mathbf{a}_p$ other than identifying all
components making up $\mathbf{v}_p$ and inferring the acceleration
on a case by case basis.
Part of $\mathbf{v}_p$ is due to our motion around the Milky Way
at 217 Km/s and a radius $r$ of about 26,000 ly. From these figures
we infer
\be
a={v^2/r}\approx 1.9\times 10^{-10} {\rm m\ s^{-2}},
\ee
which is comparable to $a_0$. In addition there are ``non-linear'' peculiar
velocity components, such as the movement of the Milky Way about
the center of the local group and the motion of the local group
toward the great attractor. These are of the order 100-200 Km s$^{-1}$, and  deriving their associated acceleration is complex.

When all these non-linear components are added, the total
points in the opposite direction to the CMB dipole (which implies
a speed of roughly 300 Km s$^{-1}$), so we may conclude that the peculiar velocity  is about
600 Km s$^{-1}$ roughly in the direction of the CMB dipole.
In cosmological linear perturbation theory there is a simple relation
between peculiar velocity and acceleration, namely
\be\label{accvel}
a_p={ \scriptstyle 3\over\scriptstyle  2}\Omega_m^{0.4}H_0 v_p,
\ee
where $H_0$ is the Hubble constant and  $\Omega_m$ is the ratio between
the matter density $\rho_m$ and the critical density $\rho_c$~\cite{peebles}.
For currently
popular values of these parameters we have $a_p\approx  1.3\times 10^{-12}  {\rm m\ s}^{-2}$. (Naively this places velocity perturbations
in the MOND regime, but the criterion for MONDian behavior
on cosmological scales should be derived from a MONDian counterpart of the above mentioned perturbation theory; this is just now becoming possible~\cite{Skordis}.)

We may guess that (\ref{accvel}) provides a good order of magnitude estimate
for the acceleration due to both linear and non-linear large-scale
perturbations. This suggests that the SS peculiar
acceleration is dominated by its motion around the center of
the galaxy. However, one cannot discount the possibility that the SS has a
significant mass in its neighborhood, e.g. undiscovered massive planets
or even a stellar companion. This would contribute to $a_p$.
In any case, this  effect has been constrained using
timing data on accurate astronomical clocks~\cite{trem}, leading to the bound
$a_p < 5\times 10^{-11} {\rm m\ s}^{-2}$.

The shift of the SP due to $\mathbf{a}_p$ is small, on the order of meters for the Earth-Sun system. But should
MONDian behavior as predicted in this paper be discovered,
the motion described by the SPs provides our best chance for a
{\it direct} measurement of the peculiar acceleration; MOND bubbles would then function as sensitive  accelerometers.

\section{Targets for LISA Pathfinder  }
\label{LPF}

As stated in the introduction, the MOND effects near the Lagrange
points are expected to be weak; however this does not mean that
they are beyond the reach of very sensitive equipment, such as that
 on board of the LISA Pathfinder   (LPF) mission~\cite{lisapf}.
Furthermore, while in transit to L1, the satellite may pass close
enough to the SP to probe the quasi-Newtonian region examined
in Section~\ref{quasi} (the extreme MONDian region described in
Section~\ref{extreme} probably requires a dedicated mission). In
the LPF mission two proof masses are suitably shielded from
radiation pressure and other annoyances that prevent testing
gravitational physics to $a_0$ accuracy in the inner parts of the
Solar system. Naturally the satellite itself has to bear radiation
pressure, but its orbit is corrected by tracking the free falling
proof masses contained in its inside. The sensitivity to tidal
stresses has been quoted as $10^{-15}\,$s$^{-2}$
(see~\cite{lisapf}).

According to Eq.~(\ref{A}),  tidal stresses at the Sun-Earth SP are of the order $A\approx 4.57\times 10^{-11}$s$^{-2}$, four orders of magnitude larger than LPF's sensitivity. The fractional corrections to Newtonian gravity contained in Eqs.~(\ref{deltaF1}) and (\ref{deltaF2}), and plotted in Fig.~\ref{corrections}, have a rough order of magnitude given by
(\ref{linpercent}). The tidal stress corresponding to $\delta F$ is thus of order $10^{-13}(r_0/r)^2\,$s$^{-2}$ for the illustrative value $k=0.03$ used in this paper. Therefore LPF would be sensitive to these MONDian
corrections if it got to within $10 r_0 \approx 3830$ Km of the
saddle. This is not overly demanding; the region
is the size of a planet.  The MOND effects may be even apparent while LPF is in transit to L1.

In contrast, the MONDian tidal stresses felt near L1 are far too
small to be at reach of this  mission.   If $r_L$ denotes L1's distance from the Sun, L1 lies   at $R- r_L\approx 1.5\times 10^6$
Km from Earth; the saddle of the Sun-Earth potential is at
$R-r_s\approx 2.6\times 10^5$ Km from Earth (see Eq.~(\ref{rs})).
Therefore L1 is $\Delta r = r_s-r_L\approx 1.24\times 10^6$~Km away from
the saddle, implying suppression of corrections to Newtonian gravity by a factor
${k\over 4\pi}{\left(r_0\over \Delta r\right)}^2\approx 2.4\times
10^{-10}$. By way of contrast, the Newtonian tidal stresses at L1 are, say for the
radial component,
 \be
 \label{numer}
 {\partial F^{(N)}_r\over
\partial r} \approx 8 \omega_E^2\approx 3.17\times
10^{-13}\,{\rm s}^{-2}
\ee
with $\omega_E$ being the angular frequency of the Earth's orbit.   This is only 2 orders of magnitude above experimental sensitivity, and so the MONDian corrections to stresses in the vicinity of L1
are 8 orders of magnitude too small for the quoted instrumental
sensitivity.

However, ``indirect'' effects may possibly  be detectable
by LPF:  effects not on its accelerometers but on its path (this
comment may apply to other L1 missions). Indeed MOND introduces a
small shift to the location of L1 and its surrounding orbits.
Combining Eqs.~(\ref{deltaF1}), (\ref{deltaF2}) and (\ref{sol1}),
we obtain an extra acceleration at L1 with radial component of signed magnitude
\be\label{deltaFL1} \delta F={4\pi\over k}{r_0\over \Delta
r}\left[{1\over 2}-{\pi\over
3\sqrt{3}}\right]a_0\approx -1.3\times 10^{-12}\,{\rm m\, s}^{-2} \,.
\ee
Hence this extra acceleration predicted by MOND points towards the SP, i.e. away from the Sun and toward the Earth. In the usual calculation, the centrifugal acceleration at L1 is exactly balanced by the gravitational one
$F^{(N)}$. This last has absolute magnitude $\omega^2_E\, r_L$ and  points away from the Sun; thus $F^{(N)}$ has to point towards it.  This is why L1 is closer to the Sun than the SP of the potential.   With the extra force (\ref{deltaFL1}) to balance, L1 is further shifted toward the Sun.  In view of Eq.~(\ref{numer}), the predicted shift is approximately 4 m.

There is a similar order of magnitude effect on the \emph{orbits} about L1, and while this is not the primary purpose of the LPF mission, we suggest that a careful monitoring of the spacecraft trajectory may be of interest to gravitational physics.

\section{The view inside the space capsule}
\label{capsule}

The form of the field $\phi$ due to the Sun and Earth as discussed in previous sections is relevant for computing the orbit of a spacecraft or the effects of tidal stresses on experiments within it.  Other questions are germane if we are interested in gravitational fields created by the spacecraft's components, as in a space reenactment of the Cavendish experiment.   These are the subject of the present section.

As long as the spacecraft's propulsion is off, it will move on a geodesic of the physical metric $\tilde g_{\alpha\beta}$ of T$e$V$e$S basically because the energy momentum tensor of matter is conserved with respect to that metric.  If we ignore relativistic corrections, this path corresponds to a Newtonian trajectory in the potential $\Phi_N+\phi$ created by the SS.   What we wish to ask is, in such an orbit what fields and forces exist within the spacecraft?

If we proceed by analogy with general relativity, one might guess that in the spacecraft's frame of reference the sum of the perturbations $\delta\Phi_N$ and $\delta\phi$ to $\Phi_N$ and $\phi$, respectively, sourced by the spacecraft's structures and free proof masses would constitute the effective gravitational potential determining relative accelerations, etc.  Below we show that this is so to sufficient approximation.  In effect this result shows that T$e$V$e$S complies with the weak equivalence principle (in a freely falling frame external gravity is cancelled out).  T$e$V$e$S does \emph{not} obey the strong equivalence principle.

In T$e$V$e$S the physical metric $\tilde g_{\alpha\beta}$ and Einstein metric $g_{\alpha\beta}$ are related by (our signature is $\{-,+,+,+\}$)
\begin{equation}
\tilde g_{\alpha\beta} =  e^{-2\phi} g_{\alpha\beta} -2\mathfrak{U}_\alpha
\mathfrak{U}_\beta \sinh (2\phi) ,
\label{physg}
\end{equation}
where  $\mathfrak{U}_\alpha$ is the eponymous vector field of the theory; it obeys $g_{\alpha\beta}\mathfrak{U}^\alpha\, \mathfrak{U}^\beta=-1 $.
   Let $g_{\alpha\beta}$ and $\tilde g_{\alpha\beta}$ represent the metrics generated by the SS.  In the frame of the freely moving spacecraft (supposed to be non-rotating) the physical metric induced by the SS will be of Minkowski form.  The transition from the global frame to the spacecraft frame is effected by projecting the said metrics with the help  of a suitable tetrad; its explicit form will not be needed here.  We use indeces $a,b,c,\, \cdots$ to label spacecraft frame vectors and tensors.

   Now $\phi$ of the SS is small; we regard its value within the spacecraft as a fixed number $\phi^{(0)}$.   Likewise, $\mathfrak{U}_a$ in the SS is a unit vector which points solely in the time direction (if we ignore the motion of the SS itself with respect to the Galaxy, etc.).  We thus regard  $\mathfrak{U}_a$ within the spacecraft as a fixed vector, $\mathfrak{U}^{(0)}_a$, which in general has small space components of order the craft's velocity.   We thus have for the Einstein metric within the spacecraft, to first order in $\phi_0$ and the spatial components of $\mathfrak{U}^{(0)}_a$,
   \be
 g_{ab}=\eta_{ab}+2\phi^{(0)}\,\eta_{ab}+4\phi^{(0)}\, \mathfrak{U}^{(0)}_a
 \mathfrak{U}^{(0)}_b +h_{ab}.
 \label{Einstg}
   \ee
The final term is the perturbation to the Einstein metric from the energy momentum tensor of spacecraft components.   Within the spacecraft  $h_{ab}$ is the only  part of $g_{ab}$ whose space variation is significant.

The $g_{ab}$ metric comes from  Einstein equations, as modified in T$e$V$e$S~\cite{teves}; here we work in  linear approximation.  As well known, the Einstein tensor can be linearized in terms of second derivatives of the perturbation to $\eta_{ab}$;  only $h_{ab}$ enters into it, in the form customary from general relativity,  because the rest of the terms are here regarded as constant.  Further, here as elsewhere, any raising of indeces can be done with $\eta^{ab}$ since the difference between it and $g^{ab}$ is already of first order in small quantities.    The sources of the modified Einstein equations contain scalar, vector and matter contributions.  Most are quadratic in $\phi$ and $\mathfrak{U}_a$ derivatives, and thus quadratic in the small $\delta\phi$ and $\delta\mathfrak{U}_a$ corrections to $\phi^{(0)}$ and $\mathfrak{U}_a$ produced by the spacecraft.  We can thus ignore these energy-momentum contributions.  Some further inspection reveals that two other contributions, that related to the T$e$V$e$S Lagrange multiplier $\lambda$, and that coming from the free function $F$, are likewise negligible.

We conclude that the only source of $h_{ab}$ is the matter of the spacecraft. In T$e$V$e$S apart from the usual source $T_{ab}$ there is one of form $T_{ab} (1-e^{4\phi})$.   Obviously because of the smallness of $\phi$ this last is negligible.  The linearized Einstein equations thus look like those in general relativity, and the relevant solutions for $h_{ab}$ are the familiar ones.  In particular, the temporal-temporal component is $h_{tt}=-2\delta \Phi_N$.  We  compute the physical metric in the spacecraft's frame by substituting Eq.~(\ref{Einstg}) into Eq.~(\ref{physg}) and setting $e^{-2\phi}=1-2(\phi^{(0)}+\delta\phi)+\cdots$ , $\sin(2\phi) =2(\phi^{(0))}+\delta\phi)+\cdots$ and $\mathfrak{U}_a=\mathfrak{U}^{(0)}_a+\delta\mathfrak{U}_a$.  We get
  \be
 \tilde g_{ab}=\eta_{ab}+h_{ab}-2\delta\phi\,(\eta_{ab}+2\, \mathfrak{U}^{(0)}_a
 \mathfrak{U}^{(0)}_b ) +\cdots,
 \label{newmetr}
   \ee
   where the terms omitted are of second order in the small quantities $\delta\phi$, $\delta\mathfrak{U}_a$ and $h_{ab}$.

   In calculating $\tilde g_{tt}$ we take note of the fact that $\mathfrak{U}^{(0)}_t$ differs from unity by a term of order the square of the spacecraft's velocity, which is of the same order as $\Phi_N$.  Such a correction is negligible in Eq.~(\ref{newmetr}).  We thus get   $ \tilde g_{tt}=-1-2(\delta\Phi_N+\delta\phi)$.  Accordingly, in the spacecraft's frame the physical gravitational potential equals  $\delta\Phi_N+\delta\phi$ as surmised earlier.

Now we know that $\delta\Phi_N$ comes from Poisson's equation. But how is $\delta\phi$ to be calculated?  Let us substitute $\phi=\phi^{(0)}+\delta\phi$ in the scalar equation (\ref{phieq}).  In the present calculations we shall take cognizance of the spatial gradient of $\phi^{(0)}$ and regard it as large compared to $\nabla\delta\phi$.   Linearizing the equation in derivatives of $\delta\phi$ leads to
\bea
 &\Delta\delta\phi&+2\xi H^i\, H^j \partial^2 \delta \phi /\partial x^i\, \partial x^j+\cdots = kG\tilde\rho,
 \label{modPoisson}
\\
&\mathbf{H}&\equiv (|\nabla\phi^{(0)}|^2)^{-1/2}\,\nabla\phi^{(0)},
\\
&\xi&\equiv d\ln \mu(Y)/\ln Y,
\eea
where $\Delta$ represents the Laplacian and $Y\equiv k\, l^2 |\nabla\phi^{(0)}|^2$.  In Eq.~((\ref{modPoisson}) the ellipsis denotes terms with  first derivatives of $\delta\phi$ only.  Quite in analogy with the eikonal approximation we shall ignore these last; presumably the spacecraft's small scale makes contributions containing only a first derivative of a varying quantity subdominant.

It should be evident that the unit vector $\mathbf{H}$ is antiparallel to the $\mathbf{U}$ (discussed in Secs.~\ref{formal}-\ref{extreme}) coming from the SS.   Aligning the coordinate system in the spacecraft's frame with its $x$ axis in the $\mathbf{H}$ direction (possible at a particular position in the orbit), we see that Eq.~(\ref{modPoisson}) is just a Poisson equation whose $x$ coordinate has been rescaled to $x(1+2\xi)^{-1/2}$.  Now for our model (\ref{muhere}) of $\mu(Y)$ we calculate
\be
\xi={\scriptstyle 1\over \scriptstyle 2}{1-\mu^4\over 1+\mu^4}\,.
\ee
It follows that when the spacecraft is in the deep MOND region ($\mu$ built with $\nabla \phi^{(0)}$ is small compared to unity), we find the $x$ direction is compressed by a factor 2.    This is a facet of the ``external field effect''~\cite{aqual} whereby MOND effects in a weak field systems are traded for quasi-Newtonian behavior but with rescaling in one direction.

By contrast, with the spacecraft deep in the quasi-Newtonian regime ($\mu\approx 1$), $\delta\phi$ is determined by the usual Poisson equation.  With $\delta\phi$ proportional to $\delta\Phi_N$ we only have a (small) rescaling of the effective gravitational constant, or equivalently of Milgrom's $\tilde\mu$, a point which has already been mentioned in Sec.~\ref{formal}.  Thus in a quasi-Newtonian environment, even the superweak fields originating in the spacecraft components behave in everyday (Newtonian) fashion.

\section{Conclusions}\label{concs}

Postulating dark matter or MONDifying the gravitational
interaction are conceptually conflicting ways of dealing with
several anomalous astrophysical observations. While it is possible
that these anomalies will themselves decide between the two
approaches, a ``direct detection'' would be far more convincing,
for example, ongoing dark matter searches finding
a particle with suitable cosmological and astrophysical
features~\cite{darkmatter}. In this paper we examined what might
constitute ``direct'' detection of MOND behavior. We predicted the
existence of regions displaying full MOND behavior well inside the
Solar system, specifically in bubbles surrounding the saddle
points of the gravitational potential. If abnormally high tidal
stresses are observed in these regions this would prove MOND
beyond reasonable doubt. Occasional astrophysical difficulties with
the theory (e.g. with regards to lensing~\cite{lens} or
cosmological density fluctuations~\cite{ferr,Skordis,silk}) would
no doubt mysteriously dissipate should such a discovery be made.

How general are our predictions? MOND's solid requirement is that $\tilde\mu(x)$ must
approach 1 as $x\gg 1$ and $x$ as $x\ll 1$; the interpolating
regime between these two asymptotic requirements is far less
constrained. In the present work this intermediate regime translates into the quasi-Newtonian calculations presented in
Sec.~\ref{quasi}. For these we chose a reasonable form for
$\mu(x)$, Eq.~(\ref{muhere}),  but we should stress that the details are model
dependent. For instance, in (\ref{mubar}) the leading correction
could have been quartic in $a_0/F$ instead of quadratic, resulting in a different power in the denominator of (\ref{curl2}). The extra
force $\delta{\mathbf F}$ would then fall off more
steeply with $r$.

Accordingly, our calculations in the quasi-Newtonian domain are simply
illustrative.  We defer to a future publication a thorough
study of the effect of the choice of  $\mu$ (as dictated by theoretical requirements and extant observations) on planetary orbits~\cite{planets}, Lagrange points, and the Pioneer anomaly~\cite{pioneer}.   By contrast our predictions for the interior of the ellipsoid (\ref{r0ellipse}), as presented in Section~\ref{extreme}, are robust predictions of the MOND scenario, and of wider validity.

We thus face a dilemma.   The strongest MOND effect and the theoretically more robust prediction is that made in Section~\ref{extreme} for the interior of the
ellipsoid (\ref{r0ellipse}). However locating it in space may be
taxing, particularly since this bubble is non-inertial. In contrast,
the quasi-Newtonian predictions, e.g.  what LISA Pathfinder
might find in the vicinity of L1, are geographically less
demanding, but the predicted effects are weaker and theoretically
less discriminative.  Thus observing what we predicted in
Section~\ref{quasi} would support the specific model (\ref{muhere}) there; however failure to observe it would hardly disprove MOND in general. The interior of the ellipsoid (\ref{r0ellipse}) is therefore the prime experimental target for a conclusive test. But one
should not despair: systems other than those examined here may
naturally reveal the inner core derived in Section~\ref{extreme}.
For example the movement of the saddle point through a diffuse medium -- say the rings of Saturn -- could be observable.

On a technical note, our calculations once more underline the limitations of the
usual folklore that MOND can be obtained by  ``taking the square
root of the Newtonian field when $\nabla\Phi\ll a_0$''. This is
only true under strict spherical symmetry; in general one must add
to the Newtonian field a curl term, which acts as a sort of
``magnetic'' gravitational field. As we have seen in the study of
saddles, this field cannot be neglected even in the quasi-Newtonian
region; and in the deep MOND region neglect of the curl component introduces downright errors in the quantitative details.  This should
serve as a warning when making naive comparisons between
theory and observation, for example in the context of clusters or
satellite galaxies.

There are other regions in the solar system where gradients of the Newtonian potential will be low, e.g at the center of near-spherical
objects. However these are obviously inaccessible. By focusing on
the saddle points of the gravitational potential in the solar system we believe we have exposed the best candidates for a direct detection of strong MONDian behavior in our own backyard.

{\bf Acknowledgments} We are grateful to Peter Coles, Ofer Lahav, Moti Milgrom, John Moffat, Norman Murray and Tim Sumners for discussions
concerning this paper. JB thanks Imperial College London and University College London, while JM thanks the University of New South Wales for hospitality.

\end{document}